\documentclass[twocolumn,showpacs,prl]{revtex4}
\begin{document}
\title{Electronic Control of Spin Alignment in $\pi$-Conjugated Molecular Magnets}
\author{Ping Huai}
\author{Yukihiro Shimoi}
\author{Shuji Abe}

\affiliation{
Nanotechnology Research Institute (NRI) and  Research Consortium \\ for
Synthetic Nano-Function Materials Project (SYNAF), National Institute of
Advanced Industrial Science and Technology (AIST), 1-1-1 Umezono, Tsukuba,
Ibaraki 305-8568, Japan}

\date{\today}

\begin{abstract}
Intramolecular spin alignment in $\pi$-conjugated molecules is studied theoretically in a model of
a Peierls-Hubbard chain coupled with two localized spins. By means of the exact diagonalization technique,
we demonstrate that a spin singlet ($S=0$) to quartet ($S=3/2$) transition can be induced by electronic
doping, depending on the chain length, the positions of the localized spins, and the sign of the
electron-spin coupling. The calculated results provides a theoretical basis for
understanding the mechanism of spin alignment recently observed in a diradical donor molecule. 
\end{abstract}

% 75.50.Xx Molecular magnets 
% 71.10.Fd Lattice fermion models (Hubbard model, etc.)
% 75.20.Hr Local moment in compounds and alloys; Kondo effect, valence fluctuations, heavy fermions 

\maketitle

Molecular magnetism has captured a large number of experimental and theoretical investigations
for decades \cite{itoh,wasserman,rajca,miller,miller2,itoh2}.
Recently Izuoka and his coworkers have succeeded in controlling intramolecular spin alignment
by charge doping in a newly designed organic molecule, thianthrene bis(nitronyl nitroxide) \cite{izuoka1}.
This molecule, consisting of $\pi$-conjugated moiety and two stable radicals, is spin singlet ($S=0$)
in its ground state and becomes spin quartet ($S=3/2$) upon one-electron oxidation, \textit{i.e.}
one-hole doping. The control of spin alignment by photoexcitation has also been demonstrated in a similar
type of molecules \cite{teki2}. The observation of such controllable intramolecular spin alignment
by external stimuli offers us very alluring prospect for its future application on ultrafast
light-switching or magnetic storage, because of flexibility and variety in designing $\pi$-conjugated molecules.  

The dominant mechanism of spin alignment in the ground state of magnetic molecules
has been well established as a topological rule \cite{mataga,ovchinnikov,lieb},
based on the dynamical antiferromagnetic spin polarization effect of $\pi$ electrons
with on-site Coulomb repulsion. However, the topological rule cannot be applied to the doped or
excited molecules, and most discussions on the spin alignment control so far have been given on the basis
of individual molecular orbitals \cite{sakurai,teki2,matsushita,yamanaka,mizouchi}.
The purpose of the present paper, therefore, is to provide a theoretical basis for
the spin alignment control by charge doping from a more general point of view. 
The central question is how the one-electron addition to or removal from the interacting
$\pi$-electron system alters the correlation between the two separate localized spins. 

For this purpose, we design a theoretical model to incorporate
delocalized $\pi$-electrons and localized spins: 
a Peierls-Hubbard model on a $N$-site linear chain with two localized spins interacting with the 
$\pi$ electrons in the form of the Kondo coupling.
The linear chain part corresponding to polyene is subject to electron-lattice interactions, 
which are taken into account in the form of the Su-Schrieff-Hegger (SSH) coupling \cite{ssh}. 
The localized spins correspond to the unpaired electrons of stable radical groups 
attached to the main chain, as realized in some substituted polyacetylenes \cite{ferritto,iwamura}.

Thus, our model is presented in the Hamiltonian,
\begin{eqnarray}
H &=& -\sum_{i,s} t_i (C^{\dagger}_{i,s}C_{i+1,s} + h.c. )  \nonumber \\
  & &  + \sum_{i} U n_{i,\uparrow} n_{i,\downarrow} 
    + \sum_{i} \frac{K}{2} (q_i -q_{i+1})^2  \nonumber \\
  & & - \frac{J}{2} (C^{\dagger}_{i1} \mathbf{\sigma} C_{i1} \cdot  \mathbf{S}_{\text{T1}}+
    C^{\dagger}_{i2} \mathbf{\sigma} C_{i2} \cdot  \mathbf{S}_{\text{T2}})~,   \\
t_i &\equiv& t_0 + \alpha(q_i-q_{i+1})~,    
\end{eqnarray}
where $C^{\dagger}_{i,s}$($C_{i,s}$) creates (annihilates) an electron with spin 
$s$($=\uparrow$ or $\downarrow$) on site $i$. The nearest-neighbor transfer integral $t_i$ depends 
on the lattice deformation with the SSH coupling constant $\alpha$.
$q_i$ is the displacement of site $i$. %from its equidistant position. 
$U$ is the on-site Coulomb repulsion energy and $n_{i,s}\equiv C^{\dagger}_{i,s}C_{i,s}$. 
$K$ is the elastic constant of the lattice, whose kinetic energy is neglected.
We define a dimensionless coupling constant $\lambda \equiv \alpha^2/(t_0 K)$ 
and dimensionless bond-lengths $\Delta_i \equiv  \alpha(q_i- q_{i+1})/t_0$.
The fourth term in the r.h.s. of Eq. (1) is the exchange coupling of two
localized $1/2$ spins ($\mathbf{S}_{\text{T1}}$ and
$\mathbf{S}_{\text{T2}}$) to the spins of $\pi$ electrons at site $i1$ and $i2$, respectively.
$C^{\dagger}_{i}$ is defined as $(C^{\dagger}_{i,\uparrow},C^{\dagger}_{i,\downarrow})$ and 
$\mathbf{\sigma}\equiv (\sigma_x,\sigma_y,\sigma_z)$ are the Pauli matrices.

We exactly diagonalized the electronic part of the Hamiltonian in Eq. (1) by the Lanczos algorithm. 
Then the lattice is treated classically and optimized by means of the Hellmann-Feynman force equilibrium
condition $\partial \left <  H \right > / \partial \Delta_i = \left <  \partial H / \partial \Delta_i  \right > = 0$
at zero  temperature, resulting in
\begin{eqnarray}
\Delta_i = \lambda \left <  \sum_{s} (C^{\dagger}_{i,s}C_{i+1,s} + h.c. ) \right >
\end{eqnarray}
where $<\cdots>$ denotes the ground state expectation value.

The alignment between the localized spins and the $\pi$ spins depends on the strength of the interaction
parameters. Typical results are shown below with the following parameters in the unit of $t_0$
(on the order of eV):
$U=5.0$, $|J|=1.0$ while $\lambda$ is allowed to vary from 0 to 0.5. 
The sign of $J$ is taken to be positive (ferromagnetic)
for most of the discussions hereafter, unless  
the antiferromagnetic (AF) coupling $J<0$ is specified for comparison.                                             
The origin of positive $J$ has been discussed on the basis of
molecular orbital calculations \cite{sakurai}.

Moreover spin alignment also relies on the length of the chain and the location of sites 
to which the localized spins are coupled. We calculated 
spin correlation function, total spin, and spin gap of the ground state
for various chain lengths and spin locations in order 
to shed light on the spin alignment of molecular systems with different structures.
In our model, the open boundary condition is imposed on the $\pi$-conjugated system.
Since our model has the electron-hole symmetry, electron doping gives the same results as
hole doping. Thus we show the results for hole doping in the present paper.

First we consider the case of the two localized spins coupled to electron spins at both ends of the chain,
\textit{i.e.} $i1=1$ and $i2=N$. The spin correlation and the lattice
deformation are plotted in Fig. 1 for both the half-filled case ($N_e=N$)
and the single hole doped case ($N_e =N-1$) with $N=10$ and $\lambda=0.25$ 
(a typical value for polyacetylene).

Figure 1(a) shows the correlations of the localized spin $\mathbf{S}_{\text{T1}}$ with
$\pi$ electron spins and $\mathbf{S}_{\text{T2}}$. 
In the case of the half-filled system, the localized spins $\mathbf{S}_{\text{T1}}$
and $\mathbf{S}_{\text{T2}}$ exhibit an antiferromagnetic correlation mediated by the chain with
the total spin $S=0$. It is straightforward to understand this antiferromagnetic spin alignment
for the half-filled ground state in terms of the topological rule. 
Because of antiferromagnetic correlation within the chain, any couple of electron spins tend to
have antiparallel alignment if there are an even number of sites between them.  
This spin correlation in the $\pi$ conjugated system together with the ferromagnetic exchange coupling
$J$ gives rise to the antiferromagnetic alignment between the localized spins.

However, when a hole is doped into the $\pi$ electron system, the correlation between
the localized spins becomes positive implying  a ferromagnetic alignment.
The spin correlation pattern in the right half of the chain ($i \geq 6$) is reversed by
hole doping leading to the parallel alignment of the localized spins.
Furthermore, the doped system turns out to be a spin quartet ($S=3/2$). 
The inset in Fig. 1(a) illustrates the alteration of spin alignment  from the spin singlet to
the spin quartet by such hole doping. 
This type of alteration corresponds to the one observed in the aforementioned
thianthrene derivative \cite{izuoka1}.
In the case of AF coupling ($J<0$), single hole doping results in
similar ferromagnetic alignment between the localized spins but with the total spin $S=1/2$.

The calculated lattice deformation patterns in Fig. 1(b) 
show that bond alternation appears in the half-filled system,
while in the case of hole doping, bond alternation becomes weak
especially around the middle of the lattice. 

The electron-lattice interaction may be a two-edged blade to the spin alignment.
On one hand, it can reduce the spin correlation in the $\pi$ electron system because 
singlet pairs of electrons are formed in the dimerized lattice. 
On the other hand, the dimerization opens a Peierls charge gap to stabilize the system against
the thermal excitation of $\pi$ electrons which may alter the spin structure.

We have examined the effect of electron-lattice coupling 
in the range of $\lambda$ = 0.0 - 0.5. The spin correlation and the gap energy for both the half-filled
case and the doped case are plotted as functions of  $\lambda$ in Fig. 2. 
The absolute value of spin correlation actually increases as $\lambda$ increases
for the half-filled ground state. For the doped case, the spin correlation is barely affected by $\lambda$.

Figure 2(b) shows the $\lambda$ dependence of the gap energy, and an inset indicates the total spin
of the first excited state as well as that of the ground state.
The gap between the quartet ground state and the doublet excited state
in the doped case is slightly enhanced by the electron-lattice coupling, while 
in the half-filled case, the gap between the singlet ground state and
the triplet excited state gradually decreases with increasing $\lambda.$
Therefore these results show that weak electron-lattice coupling is preferable in
the present case to achieve stable spin alignment control by charge doping.

As is well known, `$\pi$ topology' is of great importance to understand the spin alignment of
organic molecular system. In the present model, it includes two major factors:
the size of molecules and the positions of localized spins. 
First we proceed to the size effect on the spin alignment. 
Figure 3 shows the correlations between the localized spins, the total spins, 
and the gap energies as functions of $N$ 
with the localized spins located at both ends of the chain.
For the half-filled case, the spin correlation follows the topological rule perfectly.
The systems with even numbers of sites are spin singlet ($S=0$)
with negative correlation, while those with odd numbers of
sites are  spin quartet ($S=3/2$) with positive correlation. Furthermore, even the hole-doped 
systems demonstrate regular behavior with respect to the number of sites,
although the topological rule is generally only valid for the ground state of the half-filled system.
Thus hole doping changes the total spin in the following way:
$S=0 \rightarrow S=3/2 $, if $N$ is even; $S=3/2 \rightarrow S=0$, if $N$ is odd,
as shown in the inset of Fig. 3(b). 
It is noted that the strength of spin correlation does not decay with increasing  the chain length,
although the gap energy gradually decreases as shown in Fig. 3(c).
In the case of  AF coupling, the total spin changes by doping as: 
$S=0 \rightarrow S=1/2 $, if $N$ is even; $S=1/2 \rightarrow S=0$, if $N$ is odd. 

Next we investigate the spin alignment in the case of the localized spins attached to inner sites of
the chain. We pick up symmetric positions of the localized spins as
$(i1,i2)=(1,N), (2,N-1), \cdots$, for $N=10$. The spin correlation and the total spin are plotted
as functions of $i1$ in Fig. 4. For the half-filled case, the spin alignment is antiferromagnetic
with $S=0$, as expected by the topological rule, since there are even numbers of sites between
the coupling sites.

However, the localized spins in the doped case show positive correlation except for the location
$(i1,i2)=(5,6)$. Furthermore, those positively correlated spins can be classified into two 
categories in terms of the total spin. The doped systems are spin quartet ($S=3/2$) 
for $i1=1~\text{or}~3$ but spin doublet ($S=1/2$) for $i1=2~\text{or}~4$.
A schematic picture in the inset of Fig. 4(b) illustrates the spin alignment for these two situations. 
These results show that the topological rule cannot be applied to the doped cases.
The molecular systems with the spin locations of $(i1,i2)=(1,10)$ or $(3,8)$ are more suitable 
for the purpose of spin alignment control since they exhibit large change in total spin by doping. 
We have found that the violation of the topological rule also in the case of AF coupling.

In the present paper we have shown the calculated results for relatively
large $U$ and $J$, but even for much smaller values of these parameters the main
results remain qualitatively similar except the site dependence shown in
Fig. 4. A thorough and detailed analysis of calculated results including such
parameter dependences will be reported elsewhere.

In summary, the spin alignment control in $\pi$-conjugated molecular magnets by charge doping
has been studied in the theoretical model of a Peierls-Hubbard chain connected with two localized spins.
It is demonstrated that one-hole doping changes the total spin of the system from $S=0$ to $S=3/2$,
if the localized spins are attached to the ends of the chain with even $N$. 
The intramolecular spin alignment is not much affected by electron-lattice coupling but
depends sensitively on the topological structure of the molecular system.
It is worth emphasizing that the topological effect in the doped case is different
from that in the half-filled case. Conjugated molecules with controlled spin alignment 
as discussed in the present paper have a possibility of replacing transition-metal ions
in `molecule-based' magnets such as Prussian-blue analogs, which exhibit photoinduced
bulk magnetism \cite{sato,kawamoto}.

\begin{acknowledgments}
Authors are very grateful to Prof. A. Izouka and Dr. T. Kawamoto for valuable discussions.
This work was partly supported by NEDO under the Nanotechnology Materials Program. 
\end{acknowledgments}

%\newpage
\begin{figure}[ht]
\caption{(a) Spin correlations $<\mathbf{S}_{\text{T1}} \cdot \mathbf{S}_i>$ and  
   (b) lattice deformations $\Delta_i$ for $N=10$ and $\lambda=0.25$. The half-filled ($N_e=N$) and 
   the single hole doped cases ($N_e=N-1$) are denoted by the dashed and solid lines, respectively. 
   Inset in (a) shows schematically the change of  spin alignment  by hole doping. }
\end{figure}

\begin{figure}[ht]
\caption{(a) Spin correlations $<\mathbf{S}_{\text{T1}} \cdot \mathbf{S}_{\text{T2}}>$ and
         (b) gap energies as functions of $\lambda$ for the half-filled ($N_e=N$; dashed line)
          and the doped ($N_e=N-1$; solid line) cases with $N=10$.
          Inset in (b): the total spins of the ground and the first excited states.        
          }
\end{figure}

\begin{figure}[ht]
\caption{(a) Spin correlations $<\mathbf{S}_{\text{T1}} \cdot \mathbf{S}_{\text{T2}}>$, 
         (b) total spins, and 
         (c) gap energies as  functions of $N$ for the half-filled ($N_e=N$; open circle)
          and the doped ($N_e=N-1$; filled circle) cases with $\lambda=0.25$.
          Inset in (b) shows schematically the spin alignment alteration by hole doping.}
\end{figure}

\begin{figure}[ht]
\caption{(a) Spin correlations $<\mathbf{S}_{\text{T1}} \cdot \mathbf{S}_{\text{T2}}>$ and 
         (b) total spins as functions of $i1$, with $i2=N+1-i1$, 
         for the half-filled ($N_e=N$; open circle) and the doped 
         ($N_e=N-1$; filled circle) cases with $N=10$ and $\lambda=0.25$.
         Inset in (b): schematic illustration of spin alignment with
         total spin $S=3/2$ or $1/2$ in the doped case.}
\end{figure}
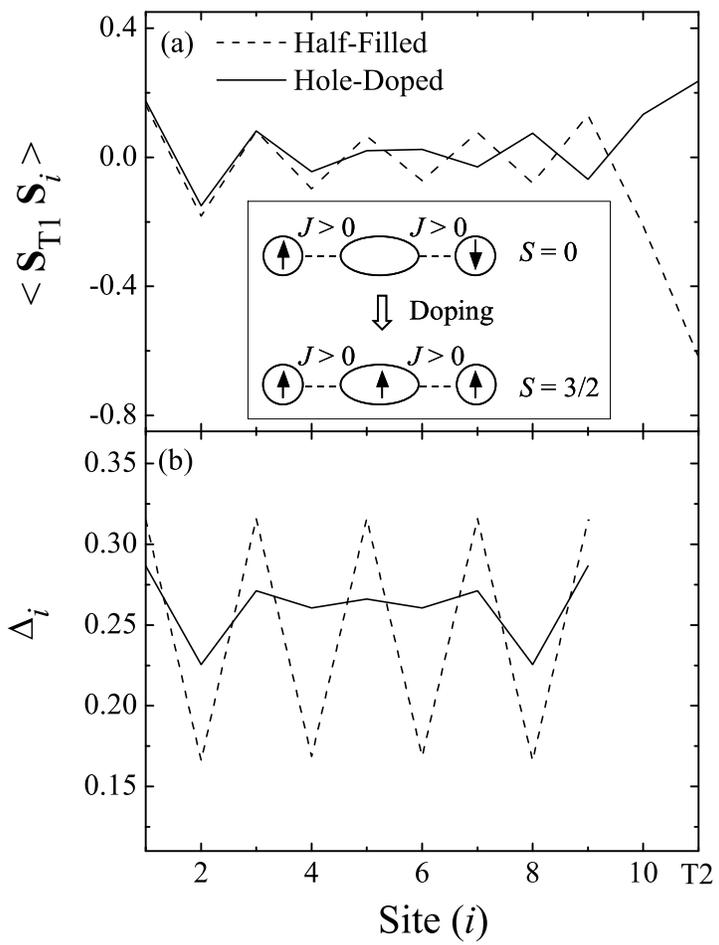
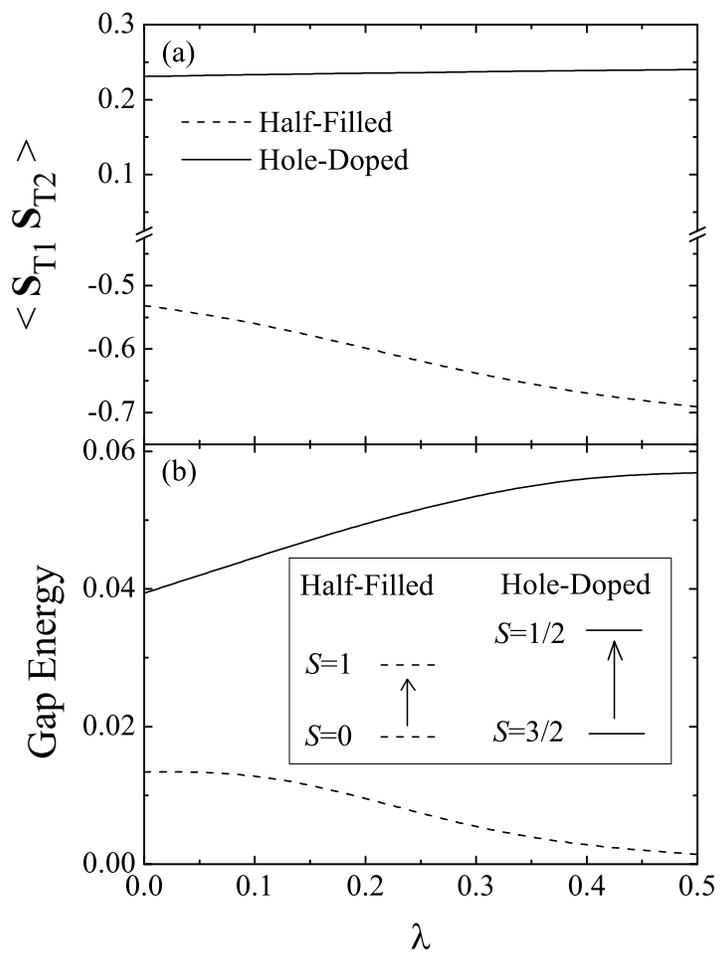
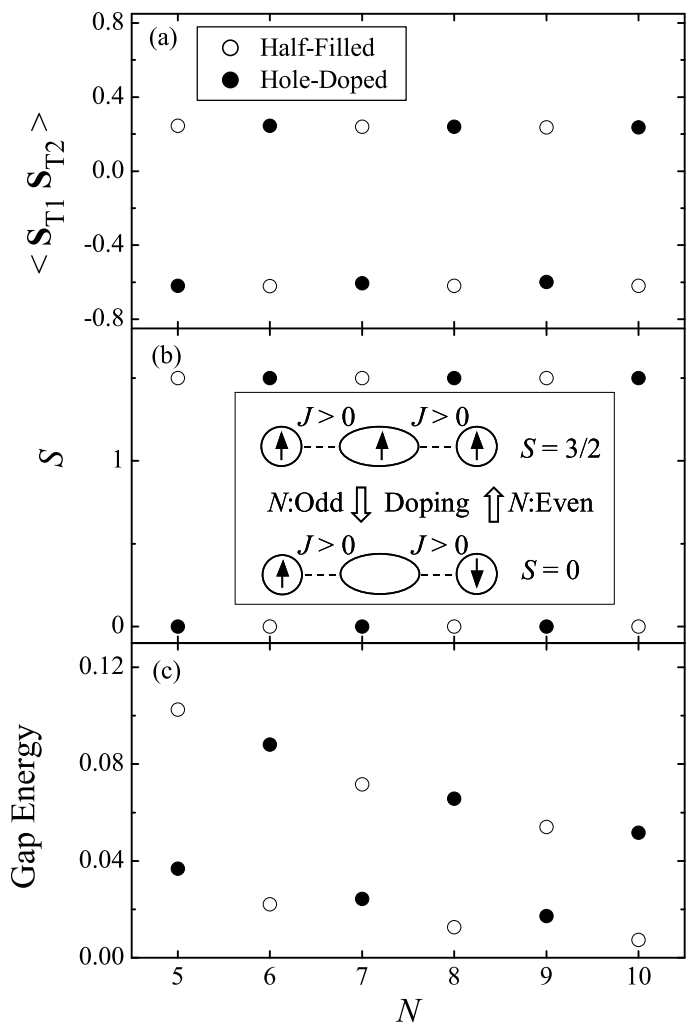
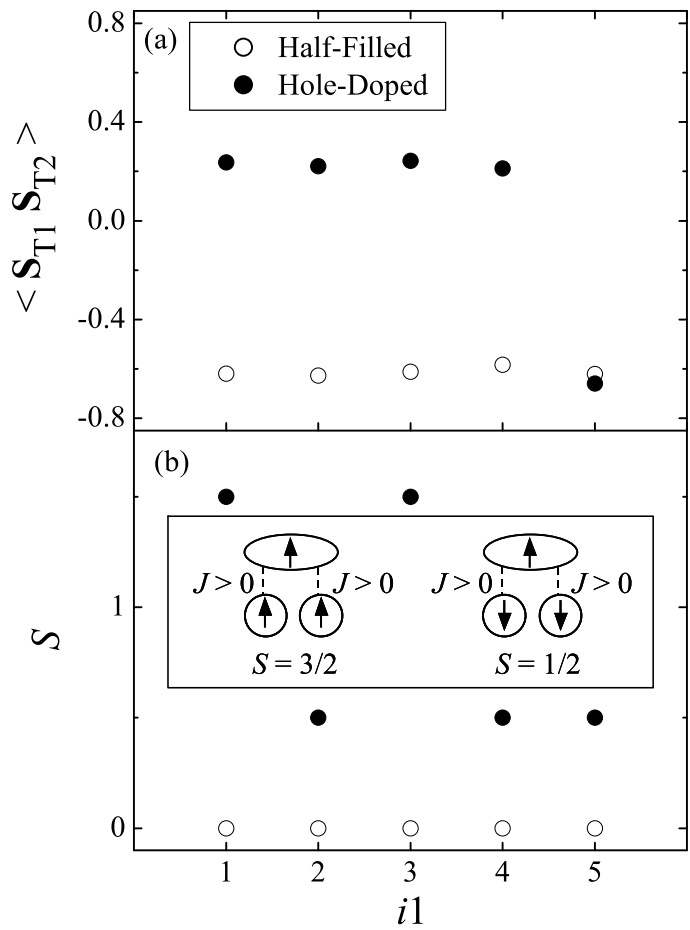
\end{document}